\documentclass[aps,twocolumn,pra,tightenlines]{revtex4}
\usepackage{amsmath,amsthm,amsfonts,amssymb,amsbsy,mathrsfs}
\usepackage{units,times,float,graphicx,bm,comment,color}
\usepackage[colorlinks,citecolor=blue,linkcolor=blue,hyperindex,dvipdfm]{hyperref}
\newcommand{\g}{\textit{g}}

\begin{document}
\title{Nonreciprocal Photon Blockade in an Asymmetric Cavity}
\author{Shao-Xiong Wu$^{1,2}$\footnote{sxwu@nuc.edu.cn}, Jin-Na Fan$^{1,2}$, Dan Yan$^2$, Cheng-Hua Bai$^2$, Qiannan Wu$^{1,2,3,4}$\footnote{qiannanwoo@nuc.edu.cn} and Mengwei Li$^{1,3,4}$\footnote{lmw@nuc.edu.cn}}
\affiliation{$^1$Shanxi Key Laboratory of Graphene Sensing Materials and Devices, North University of China, Taiyuan 030051, China\\
$^2$School of Semiconductor and Physics, North University of China, Taiyuan 030051, China\\
$^3$School of Instrument and Intelligent Future Technology, North University of China, Taiyuan 030051, China\\
$^4$Academy for Advanced Interdisciplinary Research, North University of China, Taiyuan 030051, China }
\date{\today}
\begin{abstract}
We propose a scheme to realize tunable and strong nonreciprocal photon blockade (PB) in an asymmetric Fabry-P\'{e}rot cavity. The setup consists of a single-mode optical cavity trapping a two-level atom, with the cavity coherently driven by a laser and the atom pumped by an auxiliary control field of the same frequency. By engineering quantum interference between multiple excitation pathways by adjusting the amplitude and relative phase of the control laser, we identify two distinct optimal control conditions that enable directional suppression of two-photon states. Under optimal control conditions, strong nonreciprocal PB is achieved, with a nonreciprocal ratio exceeding 30 dB over a broad operational bandwidth. The proposed protocol requires only standard coherent laser sources and is compatible with current cavity QED experimental setups, offering a practical and scalable platform for nonreciprocal quantum photonics.
\end{abstract}
\maketitle

\section{Introduction}
As a core physical resource, single-photon sources play a crucial role in both fundamental and applied research \cite{PLodahl15}, particularly with the rapid advancement of quantum optics and quantum information technologies. In fundamental research, they are extensively employed to test the foundational principles of quantum mechanics, such as quantum nonlocality \cite{NBrunner14}; in the applied domain, they are widely used in quantum communication and quantum networks \cite{FXu20}. Among various approaches to generating single-photon sources, the photon blockade (PB) effect \cite{AImamoglu97}, a typical quantum phenomenon, has attracted considerable attention due to its ability to produce high-quality single photons and its compatibility with diverse physical platforms for probing quantum nonlocality in light-matter interaction. According to the second-order photon correlation function $\g^{(2)}(0)$, photons can be classified into different states \cite{MOScully97}: the bunched state with $\g^{(2)}(0)>1$, where photons tend to arrive together, reflecting classical behavior; the anti-bunched state with $\g^{(2)}(0)<1$, where photons are emitted individually, signifying nonclassical quantum feature; and the coherent state with $\g^{(2)}(0)=1$, which lies at the boundary between the classical and quantum regimes. When the photons emitted from an optical cavity satisfy $\g^{(2)}(0)<1$, the system is considered to capture the PB effect and is restricted to a single-photon Fock state; the subsequent photon can only enter the cavity after the existing photon has leaked out.

Based on the underlying physical mechanisms, the PB effect can be classified into two categories: the conventional photon blockade (CPB) effect \cite{AImamoglu97,KMBirnbaum05} and the unconventional photon blockade (UPB) effect \cite{TCHLiew10,MBamba11,HFlayac17}. The CPB effect arises from strong coupling between the cavity and an atom or nonlinear medium, which induces a pronounced anharmonicity in the system's energy spectrum. When the driving field is resonant with the first excited state, higher-order excitations are suppressed due to large detuning from resonance, thus leading to the occurrence of the PB phenomenon. The CPB effect has been investigated in various platforms, such as cavity QED \cite{KMBirnbaum05}, superconducting circuits \cite{AJHoffman11}, and quantum dots \cite{AFaraon08}. In contrast, for systems with weak nonlinearity, the generation of the PB effect mainly depends on the UPB mechanism. The UPB typically involves utilizing multiple transition pathways, for instance, by coupling auxiliary cavities \cite{MBamba11}. The quantum destructive interference between different pathways strongly suppress the population of two-photon state, which leads to an anti-bunched state even in the absence of strong nonlinearity, thereby resulting in the PB effect. The UPB has been experimentally demonstrated by quantum dot QED systems \cite{HJSnijders18} and superconducting circuit QED platforms \cite{CVaneph18}.

The PB effect has attracted widespread attention and has been extensively explored across diverse platforms, including cavity optomechanical systems \cite{PRabl11,JQLiao13,DYWang20,JYang21,HXie24}, cavity QED systems \cite{JTang21,RTrivedi19,KHou19,YWLu22}, two-photon absorption systems \cite{FZou20,LJFeng24,YHZhou25}, bi-tone drive systems \cite{MLi22,YJing25}, waveguide-cavity QED systems \cite{ZGLu25,LLZheng25}, Tavis-Cummings systems \cite{BMarinelli25}, two-photon Jaynes-Cummings models \cite{HJLi24}, hybrid antiferromagnet-cavity quantum systems \cite{VFalch25}, neutral atom systems \cite{ACidrim20}, and so on. The PB phenomena with other quantum effects were also reported, such as Kerr effects \cite{DRoberts20,XHFan24,WZhang23}, non-Markovian effects \cite{HZShen24}, parametric amplification \cite{MCao25,YHao25,DWLiu23}, quantum loss \cite{BLi24,YZuo22,JTang25}, Zeeman splitting \cite{XSu24}, Floquet modulation \cite{SLi25}, and non-Hermitian anharmonicity \cite{ABenAsher23}. Two-photon blockade \cite{CHamsen17}, multi-mode PB \cite{YHZhou24}, phonon blockade \cite{CZhao20,XYYao22,YWang22,DGLai25} and magnon blockade \cite{ZXLiu19,KWu21,KWHuang24} that follow similar underlying physical laws have also been investigated.

In optical systems, non-reciprocity is a fundamental phenomenon. Optical nonreciprocal devices, exemplified by optical isolators, are essential components in optical path design, as they can protect sensitive optical devices from the influence of back-reflected noise. Currently, the development of magneto-free optical isolators constitutes a research frontier in the field of optics \cite{DLSounas17,NAEstep14}. By integrating the PB effect with non-reciprocity, a novel quantum phenomenon emerges: nonreciprocal photon blockade. Photons propagating in one direction the exhibit strong PB effect, resulting in anti-bunched statistics, whereas in the opposite direction, they display bunching behavior. Nonreciprocal PB effectively breaks time-reversal symmetry, establishing a direct coupling between the direction of photon transmission and the blockade effect.

Designing and fabricating nonreciprocal device based on the nonreciprocal PB effect enables the directional control of single-photon sources. Several representative platforms have been proposed for realizing nonreciprocal PB, including rotating whispering-gallery-mode cavities utilizing the Sagnac effect \cite{RHuang18,BLi19,YMLiu23,CGou23}, and asymmetric Fabry-P\'{e}rot cavities \cite{XXia21,XCGao23,SXWu25}. The non-reciprocity of the second-order photon correlation function can be demonstrated through the interaction between magnetically polarized atoms and a polarized Fabry-P\'{e}rot cavity \cite{PYang23}. Additionally, the nonreciprocal PB effects were also discussed in Refs. \cite{HZShen20,WSXue20,XShang21,AGraf22,DYWang23,NYuan24,WZhang,YZhou24,MAmazioug25,XDenga25}. Analogous to classical magneto-free optical isolators, single-photon isolators utilizing the PB effect as the core physical mechanism for non-magnetism hold profound implications for both fundamental research and further quantum technology applications. The experimental technique of asymmetric Fabry-P\'{e}rot cavities is well-established, and it offers a prominent advantage of immunity to mechanical movement, such as rotation. Building on previous works, we propose a scheme based on an asymmetric Fabry-P\'{e}rot cavity that traps a single atom and is driven by two co-frequency lasers: a driving laser and a control laser. The driving laser is coupled to the cavity mode to induce the PB effect, while the atom is pumped by the control laser in order to adjust the window of the PB effect. By precisely manipulating the amplitude and relative phase of the control field, we can selectively suppress two-photon state in the forward propagating direction, thereby achieving the nonreciprocal PB effect. This theoretical proposal can be realized readily using current cavity QED experimental setups.

The remainder of this paper is organized as follows. In Sec. \ref{sec2}, we introduce the theoretical model of a driven atom-cavity system subject to a control field, and analytically derive the steady-state probability amplitudes using the few-photon ansatz. In Sec. \ref{sec3}, we present a comprehensive investigation into nonreciprocal PB, identifying  optimal control conditions and demonstrating the manipulation of this effect via the control field. Finally, a conclusion is summarized in Sec. \ref{sec4}.

\section{The model and probability amplitudes}\label{sec2}
\begin{figure}[t]
\centering
\includegraphics[width=0.9\columnwidth]{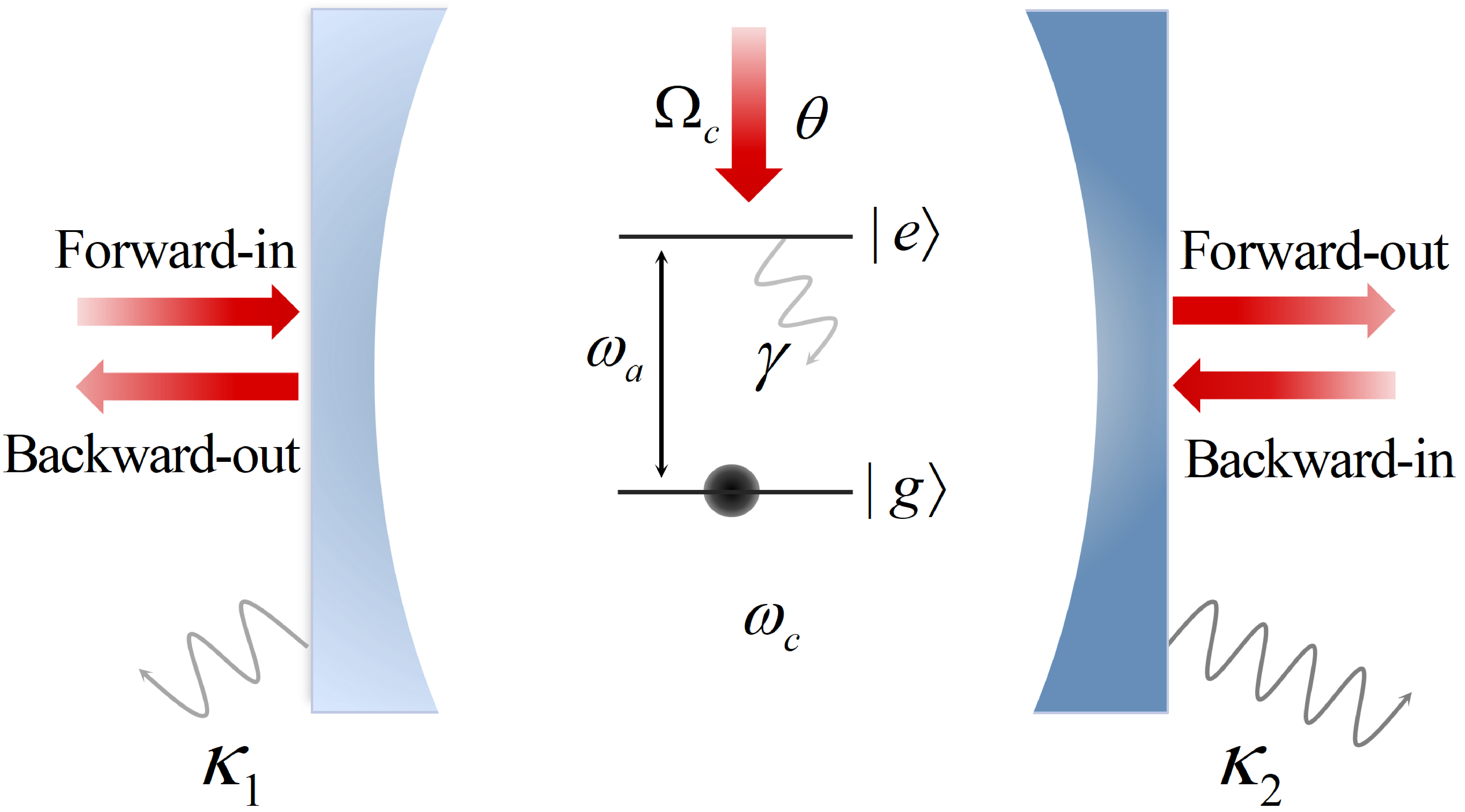}\\
\caption{Schematic of the system. A two-level atom is trapped inside an asymmetric Fabry-P\'{e}rot cavity, the cavity is driven by a coherent laser and the atom is pumped by a control laser with the same frequency. When the driving laser is injected from the left mirror toward the right, it is defined as the forward case, and the backward case corresponds to the opposite propagating direction. By adjusting the control laser, the photons exhibit photon antibunching in the forward case and bunching behaviour in the backward case, i.e., the nonreciprocal PB effect.}\label{fig1}
\end{figure}

The model of this project is schematically illustrated in Fig. \ref{fig1}. A two-level atom is trapped inside an asymmetric Fabry-P\'{e}rot cavity, and the cavity is driven by a coherent laser with amplitude $\Omega$, while the atom is simultaneously pumped by a control laser of the same frequency, characterized by amplitude $\Omega_c$ and relative phase $\theta$ with respect to the driving laser. For simplicity and without loss of generality, the phase of the driving laser is set to zero. Owing to the inherent asymmetry of the Fabry-P\'{e}rot cavity, the decay rates of the left and right mirrors are different, i.e., $\kappa_1\neq\kappa_2$, where $\kappa_1$ and $\kappa_2$ denote the photon leakage rates through the left and right mirrors, respectively. For a high-finesse cavity and neglecting the additional dissipation, the total cavity decay rate can be simplified as $\kappa=\kappa_1+\kappa_2$. The forward configuration is defined as the case where the driving laser is injected from the left mirror toward the right, and the backward configuration is defined as the reverse case. Under a fixed control parameters $\Omega_c$ and $\theta$, the system exhibits distinct photon statistics for the forward and backward directions, which enables the system to generate the nonreciprocal PB effect. The total Hamiltonian of the system (in units of $\hbar$) can be expressed as
\begin{align}
H=&\Delta_c{a^\dag}a+\Delta_a\sigma_+\sigma_- +\g(a\sigma_+ +{a^\dag}\sigma_-)\notag\\
&+\Omega(a^\dag+a) +\Omega_c(e^{i\theta}\sigma_+ +e^{-i\theta}\sigma_-),\label{Eq:H}
\end{align}
where $\Delta_c=\omega_c-\omega_d$ and $\Delta_a=\omega_a-\omega_d$ denote the detunings of the cavity mode (with resonance frequency $\omega_c$) and the atomic transition (with frequency $\omega_a$) from the driving laser frequency $\omega_d$, respectively. The operator $a(a^{\dag})$ is the annihilation (creation) operator of the cavity field, and  $\sigma_+=|e\rangle\langle\g|$ $(\sigma_-=|\g\rangle\langle e|)$ is defined as the two-level atom transition raising (lowering) operator between the excited state $|e\rangle$ and the ground state $|\g\rangle$.

To investigate the dynamics of the system, we adopt a phenomenological non-Hermitian description by incorporating the imaginary dissipation terms into the Hamiltonian (\ref{Eq:H}). The effective non-Hermitian Hamiltonian can be written as
\begin{align}
H_{\text{non}}=&\tilde{\Delta}_c{a^\dag}a+\tilde{\Delta}_a\sigma_+\sigma_- +\g(a\sigma_+ +{a^\dag}\sigma_-)\notag\\
&+\Omega(a^\dag+a) +\Omega_c(e^{i\theta}\sigma_+ +e^{-i\theta}\sigma_-),\label{Eq:Hnon}
\end{align}
where the effective detunings are defined as $\tilde{\Delta}_c=\Delta_c-i\kappa/2$ and $\tilde{\Delta}_a=\Delta_a-i\gamma/2$, with $\kappa$ and $\gamma$ denoting the cavity decay rate and the atomic spontaneous emission rate, respectively. We assume that the driving and control fields are of the same order of magnitude, i.e., $\Omega\sim\Omega_c$, which will be verified by the numerical results presented later. Under the weak-driving regime, i.e., $\{\Omega,\Omega_c\}\ll\kappa$, the system's dynamics can be well approximated within a truncated Hilbert space up to the three-excitation subspace, and the corresponding state ansatz is given by
\begin{align}
|\psi(t)\rangle=&C_{0\g}|0\g\rangle+C_{1\g}|1\g\rangle+C_{2\g}|2\g\rangle+C_{3\g}|3\g\rangle\notag\\ &+C_{0e}|0e\rangle+C_{1e}|1e\rangle+C_{2e}|2e\rangle.\label{Eq:psit}
\end{align}
The time evolution of the state $|\psi(t)\rangle$ is governed by the Schr\"{o}dinger equation under the effective non-Hermitian Hamiltonian (\ref{Eq:Hnon}), and is expressed as $i|\dot{\psi}(t)\rangle=H_{\text{non}}|\psi(t)\rangle$. Although the norm of the state (\ref{Eq:psit}), i.e., $\langle\psi(t)|\psi(t)\rangle$, is not conserved under non-Hermitian dynamics, in the steady-state limit ($t\to\infty$), one can impose the normalization condition $\langle\psi|\psi\rangle=1$ on the asymptotic state for the purpose of evaluating the second-order photon correlation function.

Under the weak-driving approximation, the ground state $|0\g\rangle$ can be assumed to be dominant and remains nearly unpopulated by excitations. Neglecting higher-order terms (e.g., states with more than three excitations), the dynamics of probability amplitudes is governed by the following set of differential equations
\begin{align}
i\dot{C}_{1\g}=&\Omega+\tilde{\Delta}_cC_{1\g}+\g{C_{0e}}+\Omega_ce^{-i\theta}C_{1e}+\sqrt{2}\Omega{C_{2\g}},\notag\\
i\dot{C}_{2\g}=&\sqrt{2}\Omega{C_{1\g}}+2\tilde{\Delta}_cC_{2\g}+\sqrt{2}\g{C_{1e}},\notag\\
i\dot{C}_{3\g}=&\sqrt{3}\Omega{C_{2\g}}+3\tilde{\Delta}_cC_{3\g}+\sqrt{3}\g{C_{2e}},\label{Eq:weifen}\\
i\dot{C}_{0e}=&\Omega_ce^{i\theta}+\g{C_{1\g}}+\tilde{\Delta}_aC_{0e}+\Omega{C_{1e}},\notag\\
i\dot{C}_{1e}=&\Omega_ce^{i\theta}C_{1\g}+\sqrt{2}\g{C_{2\g}}+\Omega{C_{0e}}+(\tilde{\Delta}_a+\tilde{\Delta}_c)C_{1e},\notag\\
i\dot{C}_{2e}=&\Omega_ce^{i\theta}C_{2\g}+\sqrt{3}\g{C_{3\g}}+\sqrt{2}\Omega{C_{1e}}+(\tilde{\Delta}_a+2\tilde{\Delta}_c)C_{2e}.\notag
\end{align}
Solving the above differential equations analytically, one can obtain explicit expressions for the steady-state probability amplitudes
\begin{align}
C_{1\g}=&\frac{\mathcal{F}_1+\mathcal{F}_2}{\mathcal{D}},\quad
C_{2\g}=\frac{\mathcal{F}_3+\mathcal{F}_4}{\sqrt{2}\mathcal{D}},\notag\\
C_{0e}=&\frac{c_{01}+c_{02}}{\mathcal{D}},\quad
C_{1e}=\frac{c_{11}+c_{12}}{\mathcal{D}},\notag\\
C_{3\g}=&\frac{c_{31}+c_{32}+c_{33}}{\sqrt{6}(\Delta_{ac}+\tilde{\Delta}_c^2)\mathcal{D}},\label{Eq:C}\\
C_{2e}=&\frac{c_{21}+c_{22}+c_{23}}{\sqrt{2}(\Delta_{ac}+\tilde{\Delta}_c^2)\mathcal{D}}.\notag
\end{align}
For notational simplicity, we introduce the following coefficients to represent the probability amplitudes in Eq. (\ref{Eq:C}): $\mathcal{D}=\g^4+(\tilde{\Delta}_c^2-\Omega^2)\Delta_{ac}'-\g^2\tilde{\Delta}_{ac} -\tilde{\Delta}_a\tilde{\Delta}_c\Omega_c^2 +2\Delta_{\g}\Omega_c\cos[\theta]$, $\mathcal{F}_1=\Omega_ce^{i\theta}\g(\Delta_{ac}+\Omega^2)$, $\mathcal{F}_2=\Omega[\g^2\tilde{\Delta}_a-\tilde{\Delta}_c(\Delta_{ac}'+\Omega_c^2)]$, $\mathcal{F}_3=\Omega_c^2e^{2i\theta}\g^2-2\Omega_ce^{i\theta}\Delta_g$, $\mathcal{F}_4=\Omega^2(\g^2+\Delta_{ac}'+\Omega_c^2)$, $c_{31}=\Omega_c^3e^{3i\theta}\g^3-3\Omega_c^2e^{2i\theta}\g^2(\tilde{\Delta}_a+2\tilde{\Delta}_c)\Omega$, $c_{32}=\Omega_ce^{i\theta}\g\Omega^2[3\g^2+3(\tilde{\Delta}_a+\tilde{\Delta}_c)(\tilde{\Delta}_a+2\tilde{\Delta}_c) -3\Omega^2+\Omega_c^2]$, $c_{33}=-\Omega^3[\g^2(3\tilde{\Delta}_a+4\tilde{\Delta}_c)+(\tilde{\Delta}_a+2\tilde{\Delta}_c)(\Delta_{ac}'+\Omega_c^2)]$, $c_{01}=\g\Omega(\Delta_{ac}+\Omega^2-\Omega_c^2)-\Omega_c^2e^{2i\theta}\g\Omega$, $c_{02}=\Omega_ce^{i\theta}[\tilde{\Delta}_a\Omega^2+\tilde{\Delta}_c(\Omega_c^2-\Delta_{ac})]$, $c_{11}=-\Delta_{\g}\Omega-\Omega_c^2e^{2i\theta}\g\tilde{\Delta}_c$, $c_{12}=\Omega_ce^{i\theta}\Omega(\Delta_{ac}+2\g^2-\Omega^2)$; $c_{21}=\Omega_c^2e^{2i\theta}\g\Omega[\g^2+2\tilde{\Delta}_c(\tilde{\Delta}_a+2\tilde{\Delta}_c)] -\Omega_c^3e^{3i\theta}\g^2\tilde{\Delta}_c$, $c_{22}=\g\Omega^3[\g^2+(\tilde{\Delta}_a+\tilde{\Delta}_c)(\tilde{\Delta}_a+2\tilde{\Delta}_c)-\Omega^2+\Omega_c^2]$, and $c_{23}=-\Omega_ce^{i\theta}\Omega^2\{\g^2(2\tilde{\Delta}_a+5\tilde{\Delta}_c) +\tilde{\Delta}_c[(\tilde{\Delta}_a+\tilde{\Delta}_c)(\tilde{\Delta}_a+2\tilde{\Delta}_c)-3\Omega^2+\Omega_c^2]\}$. The symbols $\Delta_{ac}$, $\Delta_{ac}'$, $\tilde{\Delta}_{ac}$ and $\Delta_{\g}$ in above coefficients are defined as: $\Delta_{ac}=\tilde{\Delta}_c(\tilde{\Delta}_a+\tilde{\Delta}_c)-\g^2$, $\Delta_{ac}'=\tilde{\Delta}_a(\tilde{\Delta}_a+\tilde{\Delta}_c)-\Omega^2$, $\tilde{\Delta}_{ac}=2\tilde{\Delta}_a\tilde{\Delta}_c+\tilde{\Delta}_c^2+2\Omega^2$, and $\Delta_{\g}=\g(\tilde{\Delta}_a+\tilde{\Delta}_c)\Omega$.

The hierarchy of excitation amplitudes follows from the weak-driving expansion: the single-excitation amplitudes scale as $\{C_{1\g},C_{0e}\}\sim\mathcal{O}(\Omega/\kappa)$; the two-excitation amplitudes scale as $\{C_{2\g},C_{1e}\}\sim\mathcal{O}(\Omega^2/\kappa^2)$; and the three-excitation amplitudes scale as $\{C_{3\g},C_{2e}\}\sim\mathcal{O}(\Omega^3/\kappa^3)$. The output field $b$ is related to the intracavity field $a$ via the input-output relation $b_{\text{out}}=-i\sqrt{\kappa_2}a$. Consequently, the statistics of emitted photons are determined by the second-order correlation function of the output field, which captures the same properties as those of the cavity state (\ref{Eq:psit}). Explicitly, it is given by
\begin{align}
\g^{(2)}(0)=&\frac{\langle{a}^\dag{a}^\dag{a}a\rangle}{\langle{a}^\dag{a}\rangle^2}\label{Eq:g20}\\
=&\frac{2|C_{2\g}|^2+2|C_{2e}|^2+6|C_{3\g}|^2}{(|C_{1\g}|^2+|C_{1e}|^2+2|C_{2\g}|^2+2|C_{2e}|^2+3|C_{3\g}|^2)^2}.\notag
\end{align}

\begin{figure*}
\centering
\includegraphics[width=0.75\columnwidth]{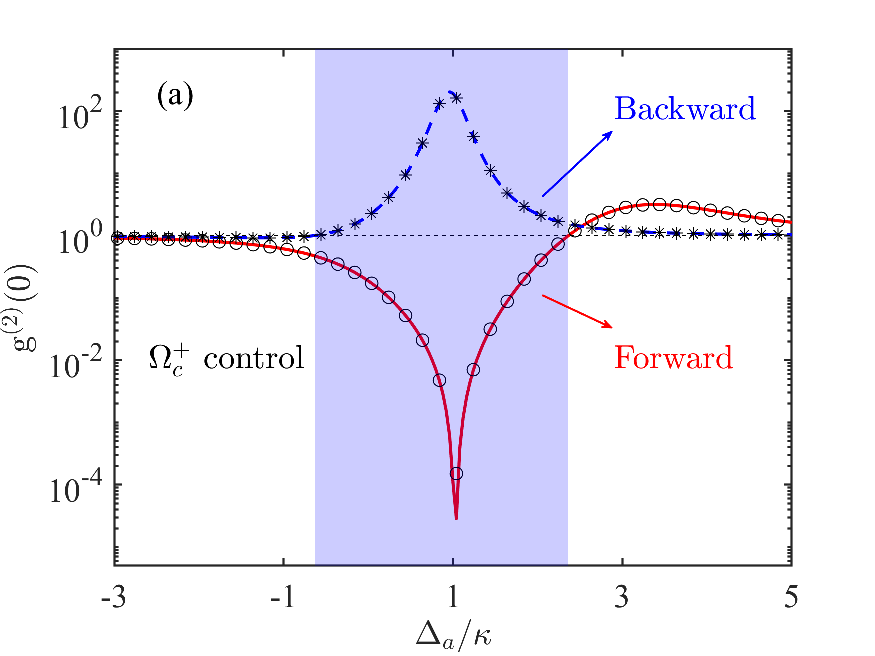}
\includegraphics[width=0.75\columnwidth]{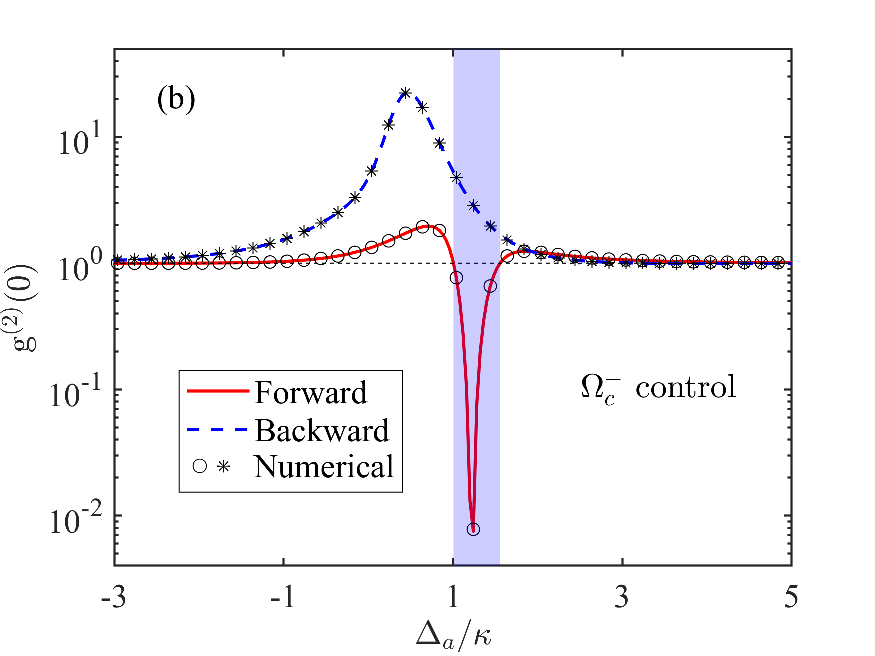}\\
\includegraphics[width=0.75\columnwidth]{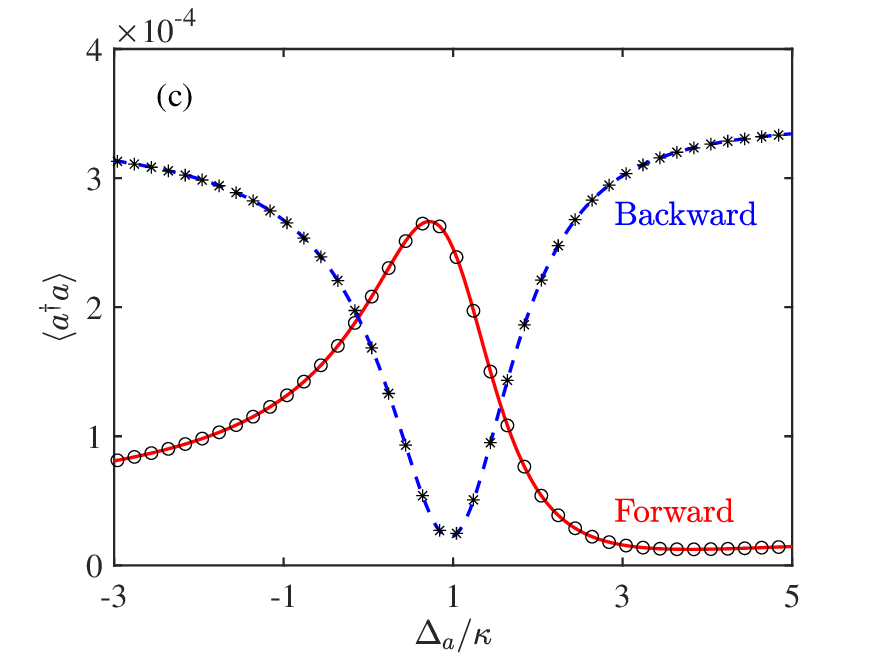}
\includegraphics[width=0.75\columnwidth]{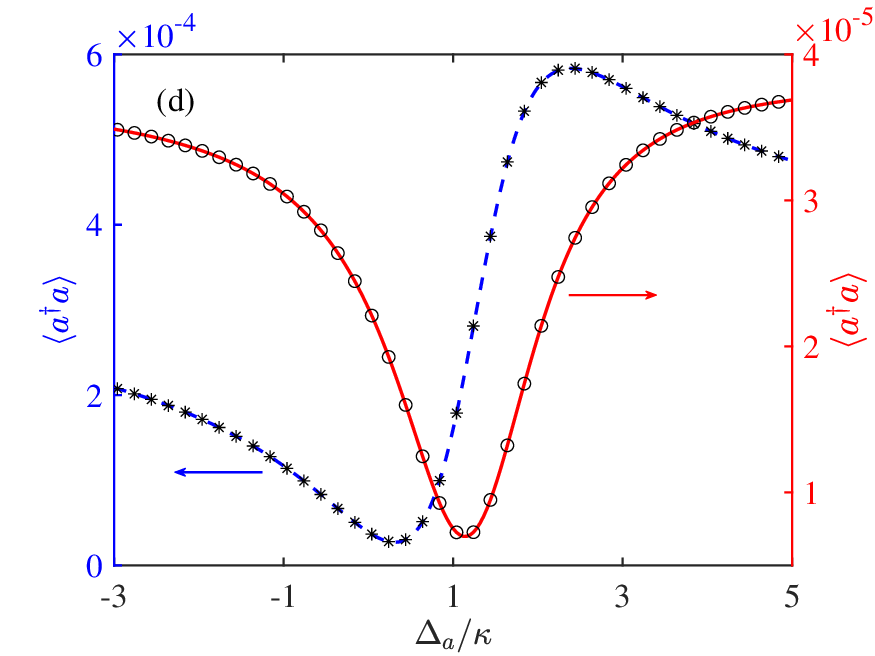}
\caption{The second-order correlation function $\g^{(2)}(0)$ and the mean intracavity photon number $\langle a^{\dagger}a\rangle$  as functions of the atomic detuning $\Delta_a$. Panels (a) and (c) correspond to the optimal control condition $\Omega_c^+$, while panels (b) and (d) represent the optimal control condition $\Omega_c^-$. Red solid lines and blue dashed lines denote the analytical results derived via the probability amplitude method for the forward and backward cases, respectively. Black circles o and asterisks $*$ mean the corresponding numerical solutions obtained from the quantum master equation.}\label{fig2}
\end{figure*}

In addition to the analytical treatment based on the probability amplitude method, the full dissipative dynamics of the system can also be consistently described within the quantum master equation, which accounts for both coherent evolution and irreversible decay processes, and be explicitly given by
\begin{align}
\dot{\rho}=-i[H,\rho]-\frac{\kappa}{2}\mathcal{L}[a]\rho -\frac{\gamma}{2}\mathcal{L}[\sigma_-]\rho.\label{Eq:master}
\end{align}
In Eq. (\ref{Eq:master}), the operator $\rho$ denotes the density matrix of the composite cavity-atom system. The dissipative dynamics are modeled using the standard Lindblad form with superoperator $\mathcal{L}[o]\rho=o^{\dag}o\rho-2o\rho{o^{\dag}}+ \rho{o^{\dag}}o$. For simplicity and without loss of generality, the system is assumed to be coupled with a zero-temperature vacuum reservoir, so that thermal excitations can be neglected.

\section{The nonreciprocal photon blockade}\label{sec3}
A perfect PB requires the suppression of the population in the second excited state, where the ideal situation is that $|C_{2\g}|=0$. The probability amplitude $C_{2\g}$ in Eq. (\ref{Eq:C}) contains a quadratic term in the control laser, i.e., a binomial function with respect to $\Omega_ce^{i\theta}$. Imposing  the condition $|C_{2\g}|=0$ to suppress the two-photon excitation yields two distinct solutions for the optimal control field, and it can be achieved by precisely manipulating the control laser, such that the optimal control condition is consistent with
\begin{align}
\frac{\Omega_ce^{i\theta}}{\Omega}=\frac{\g(\tilde{\Delta}_a+\tilde{\Delta}_c)\pm \sqrt{\g^2\Delta_{ac}-\Omega^2\Delta_{ac}'}}{\g^2+\Omega^2}.\label{Eq:optimal}
\end{align}
Due to the contribution of higher excited states, the second-order correlation function $\g^{(2)}(0)$ remains nonzero even when the optimal control condition (\ref{Eq:optimal}) is satisfied. Nevertheless, $\g^{(2)}(0)$ attains a pronounced local minimum in the vicinity of the optimal point specified by Eq. (\ref{Eq:optimal}). In contrast to Ref. \cite{SXWu25}, the optimal control condition in this study admits two distinct sets of solutions; we will analyze how these parameters affect the nonreciprocal PB effect.

The system parameters are chosen based on typical values for the cavity QED experiment setup \cite{PYang23}: $\kappa/2\pi=3.7$ MHz, $\gamma/\kappa=0.7$, $\kappa_1/\kappa=0.1$, $\g/\kappa=1.1$, and $b_{\text{in}}/\sqrt{\kappa}=0.02$. We first fix the detuning between the driving laser and the cavity field at $\Delta_c/\kappa=0.9$, and subsequently tune the frequency of the control laser to change the atomic detuning $\Delta_a$. The cavity-driving and atomic control lasers are phase-locked to a common reference with a relative phase $\theta$. In principle, varying the frequency of the control laser would shift the cavity detuning $\Delta_c$ due to the driving and control lasers are assumed to be the same frequency; however, this shift can be compensated by an active stabilization of the cavity resonance, thereby preserving the fixed atomic detuning $\Delta_c/\kappa=0.9$. The employed atomic transition corresponds to the D1 line of $^{133}$Cs atom, which has a transition wavelength of 852 nm and provides a two-level structure required for the realization of the PB effect. For a driving-field amplitude of $b_{\text{in}}/\sqrt{\kappa}=0.02$, the corresponding input laser power is about 0.3 fW, which is well within the capabilities of current experimental setups with appropriate attenuation. By manipulating the control laser, the system can be operated in a regime where perfect PB is attained.

For clarity, we denote the condition of the control laser by $\Omega_c^+$ when the positive root of Eq. (\ref{Eq:optimal}) is selected, and by $\Omega_c^-$ when the negative root is used. In Fig. \ref{fig2}, we plot both the second-order correlation function $\g^{(2)}(0)$ and the mean intracavity photon number $\langle a^{\dagger}a\rangle$ as functions of the atomic detuning $\Delta_a$. In Fig. \ref{fig2}(a), the optimal control condition is chosen based on the positive solution
\begin{align}
\Omega_c^+=\frac{\Omega[\g(\tilde{\Delta}_a+\tilde{\Delta}_c)+ \sqrt{\g^2\Delta_{ac}-\Omega^2\Delta_{ac}'}]}{\g^2+\Omega^2}.\label{Eq:+}
\end{align}
The optimal atomic detuning corresponding to the perfect PB effect is set to $\Delta_a=\text{Re}[\g^2/\tilde{\Delta}_c]$. For the forward case, one can find that the second-order correlation function becomes extremely small at the setting atomic detuning, with $\g^{(2)}(0)\sim10^{-4}\ll1$, indicating that the transmission of the two-photon state is effectively inhibited. This clearly demonstrates antibunching behavior, which signifies the realization of a strong PB effect. In contrast, when the direction of the driving laser is reversed while keeping other system parameters unchanged, the second-order correlation function $\g^{(2)}(0)$ in the same detuning region increases substantially and becomes much larger than 1, where the photons exhibit a pronounced bunching feature and reflect a strong tendency for correlated multi-photon transmission.

In light-blue shaded region of Fig. \ref{fig2}(a), the nonreciprocal PB effect is realized. The red solid line denotes the analytical solution (\ref{Eq:g20}) derived via the probability amplitude method for the forward case, while the black circles correspond to the numerical solutions obtained from the quantum master equation (\ref{Eq:master}). Similarly, the blue dashed line represents the analytical solution for the backward case, and the black asterisks indicate the associated numerical results. Excellent agreement between the analytical and numerical approaches is evident across the entire parameter range for both forward and backward cases. The black dotted line marks the coherent state, i.e., $\g^{(2)}(0)=1$. Below this line, i.e., $\g^{(2)}(0)<1$, the emitted photon exhibits antibunching, indicating that the system is in the PB regime; conversely, above this line ($\g^{(2)}(0)>1$), the photons display bunching feature, characteristic of a non-blockaded state. Crucially, within the same parameter interval, specifically the light-blue region, strong nonreciprocity emerges: the system can be switched from an antibunching and blockade regime in the forward case to a bunching and non-blockaded state in the backward case.

Fig. \ref{fig2}(c) shows the mean intracavity photon number $\langle{a^{\dag}a}\rangle$ using the same parameters as in Fig. \ref{fig2}(a). In the region where nonreciprocal PB occurs, the forward case exhibits a pronounced peak of photon population, consistent with the single-photon resonance condition at the optimal atomic detuning of the perfect PB effect. By contrast, the backward case shows a distinct dip in the same detuning region. These complementary features, simultaneous suppression of the second-order correlation statistics $\g^{(2)}(0)$ and enhancement of photon population in the forward case, alongside opposite behavior in the backward case, confirm the robust realization of a strong nonreciprocal PB effect.

\begin{figure}[t]
\centering
\includegraphics[width=0.75\columnwidth]{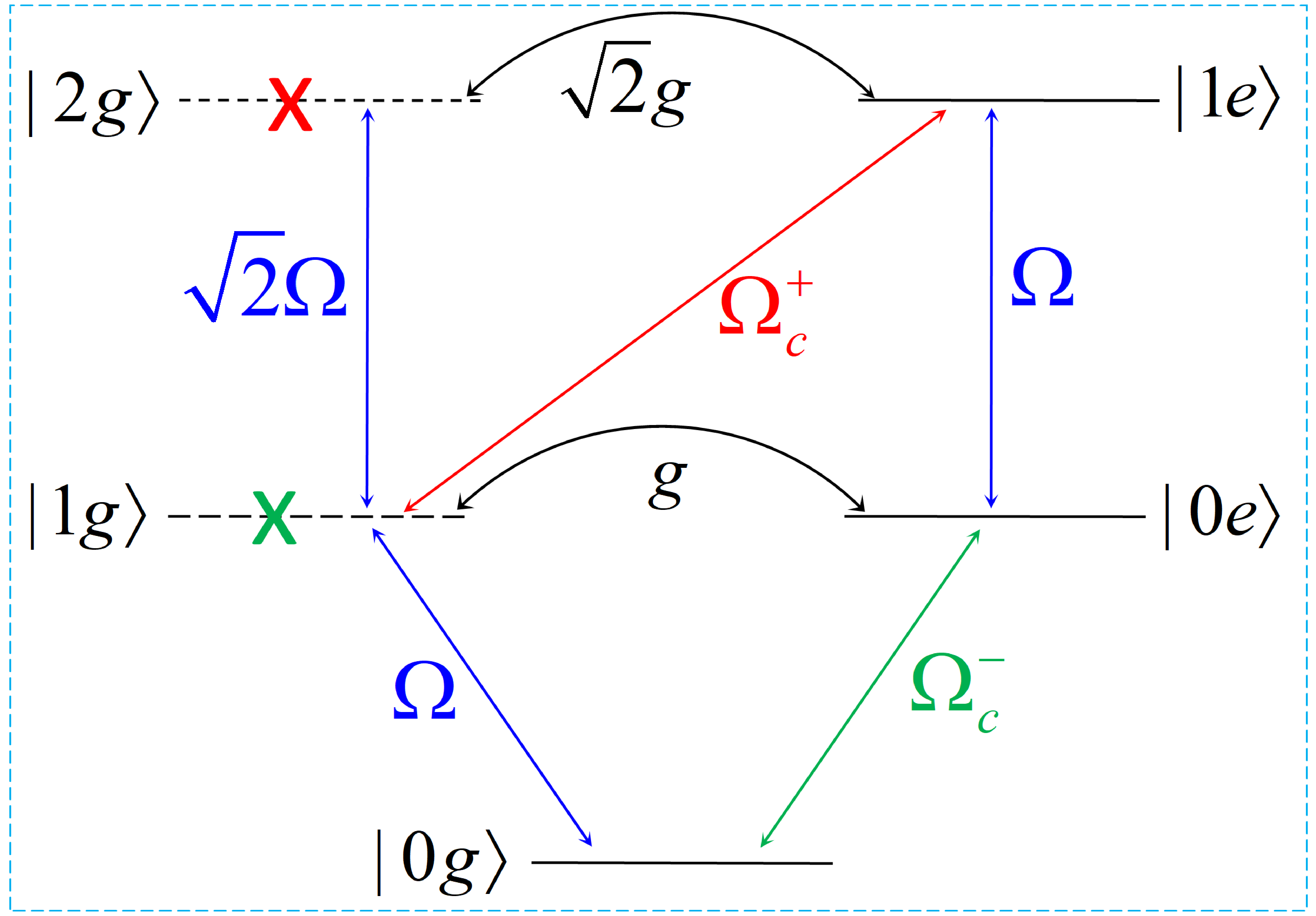}\\
\caption{Schematic of the energy-level structure and the transition pathways. The blue arrows denote transitions induced by the driving field, while the red/green arrow represents the pumping by the control condition $\Omega_c^+$ or $\Omega_c^-$.} \label{fig3}
\end{figure}

The property of the system can also be analyzed under the optimal control condition related to  the negative root of Eq. (\ref{Eq:optimal})
\begin{align}
\Omega_c^-=\frac{\Omega[\g(\tilde{\Delta}_a+\tilde{\Delta}_c)- \sqrt{\g^2\Delta_{ac}-\Omega^2\Delta_{ac}'}]}{\g^2+\Omega^2}.\label{Eq:-}
\end{align}
In Figs. \ref{fig2}(b) and \ref{fig2}(d), we present the second-order correlation function $\g^{(2)}(0)$ and the intracavity photon number $\langle{a^{\dag}a}\rangle$, respectively, evaluated under the optimal control condition $\Omega_c^-$. In comparison with the results obtained using the condition $\Omega_c^+$, the result utilizing the condition $\Omega_c^-$ exhibits a weaker PB phenomenon, in the scale of $\g^{(2)}(0)\sim10^{-2}$, and simultaneously yields a lower cavity photon population in the region where PB occurs. Moreover, the bandwidth of the nonreciprocal PB regime is noticeably narrower. These obvious differences between the two optimal control conditions can be explained by the energy-level transition diagram shown in Fig. \ref{fig3}, which highlights the distinct quantum destructive interference pathways associated with each control condition.

The optimal control conditions, denoted by $\Omega_c^+$ (\ref{Eq:+}) and $\Omega_c^-$ (\ref{Eq:-}), are necessary and essential to realize the nonreciprocal PB effect, since $\g^{(2)}(0)$ in Eq. (\ref{Eq:g20}) is approximatively unrelated to the amplitude of the driving laser $\Omega$ when the control field is absent. When the optimal control field $\Omega_c^+$ (the red arrow in Fig. \ref{fig3}) is applied, the dominant quantum destructive interference occurs between the transition pathways $|1\g\rangle\to|2\g\rangle$ and $|1\g\rangle\to|1e\rangle\to|2\g\rangle$. By appropriately tuning both the amplitude and relative phase of the control field $\Omega_c^+$, these two pathways can be brought into considerable destructive interference, effectively inhibiting the population of state $|2\g\rangle$ and yielding a strong PB effect. In contrast, the mechanism of PB under the optimal control field $\Omega_c^-$ (the green arrow in Fig. \ref{fig3}) lies so close to the condition $|C_{1\g}|=0$. With the weak-driving approximation and neglecting high-order terms $\Omega^2$ and $\Omega_c^2$, the solution of $|C_{1\g}|=0$ can be reduced to $\Omega_ce^{i\theta}=\tilde{\Delta}_a\Omega/\g$. It can form a quantum destructive interference between pathways $|0\g\rangle\to|1\g\rangle$ and $|0\g\rangle\to|0e\rangle\to|1\g\rangle$, the population of state $|1\g\rangle$ is suppressed, and the direct excitation from state $|1\g\rangle$ to state $|2\g\rangle$ is consumingly suppressed. Nevertheless, another two-photon excitation process can still be simulated via the cascade path $|0e\rangle\to|1e\rangle\to|2\g\rangle$. Consequently, even though the PB effect can be realized, it is very weak and the second-order correlation function $\g^{(2)}(0)$ reaches only $\sim{10^{-2}}$ as shown in Fig. \ref{fig2}(b), which is significantly weaker than that under the condition $\Omega_c^+$. Simultaneously, the mean intracavity photon number is substantially reduced, as illustrated in Fig. \ref{fig2}(d). The above analysis applies to the forward configuration case. When the direction of the driving laser is reversed while keeping the parameters of control field unchanged, the quantum destructive interference condition required for the PB effect is no longer satisfied, and the second-order correlation $\g^{(2)}(0)$ will rise significantly, which indicates the absence of the PB effect for backward case. This directional-dependent asymmetry in quantum interferences constitute the physical mechanism of the nonreciprocal PB effect.

\begin{figure}[t]
\centering
\includegraphics[width=0.75\columnwidth]{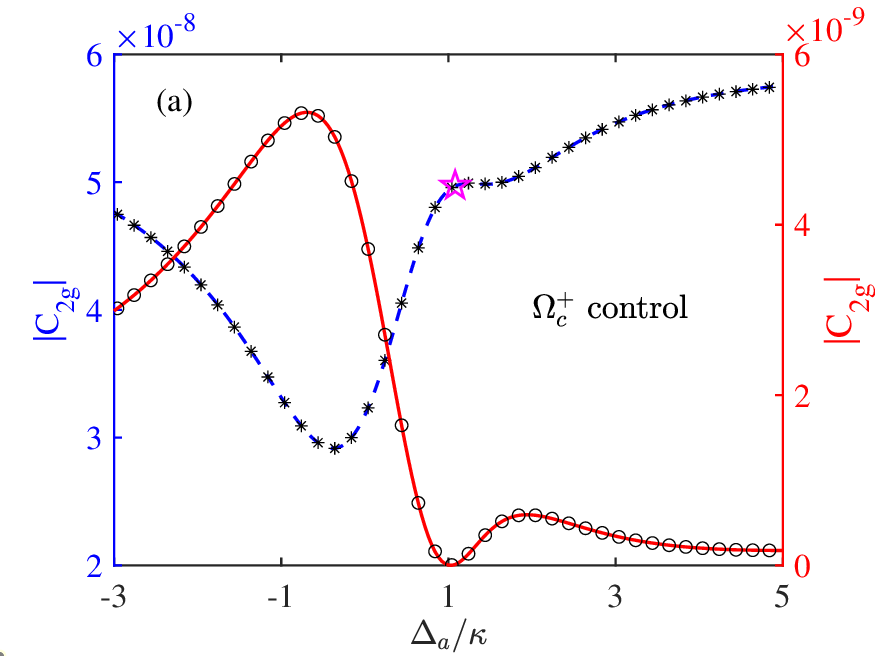}\\
\includegraphics[width=0.75\columnwidth]{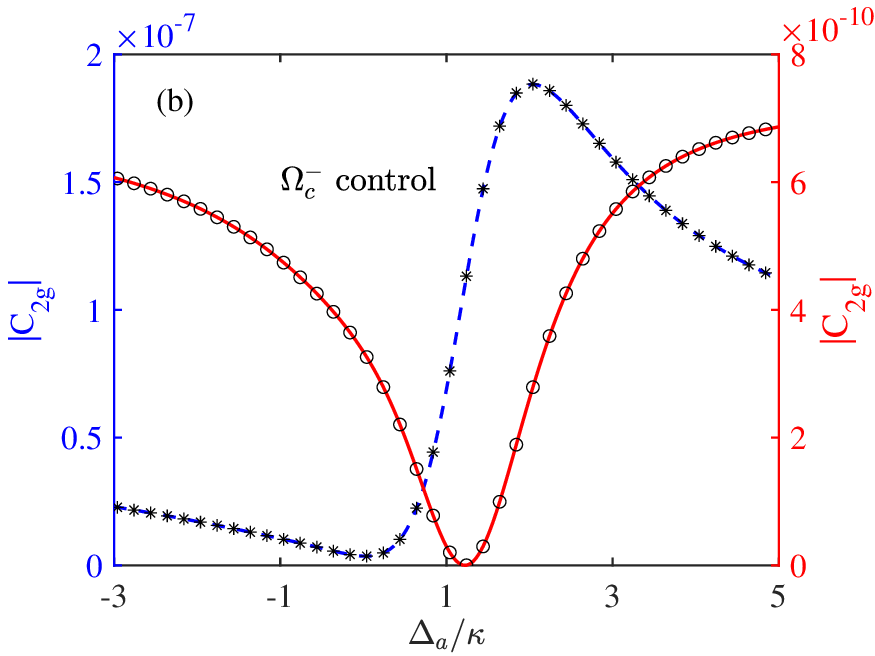}\\
\caption{The population of two-photon state $|2\g\rangle$ as a function of the atomic detuning $\Delta_a$. Panel (a) corresponds to the optimal control condition $\Omega_c^+$, and panel (b) to $\Omega_c^-$ control. The parameters are chosen those in Fig. \ref{fig2}.} \label{fig4}
\end{figure}

In Fig. \ref{fig4}, we further present the population of the second excited state $|2\g\rangle$. When the optimal control condition $\Omega_c^+$ is employed, the population of $|2\g\rangle$ exhibits a pronounced decrease in the optimal atomic detuning of the perfect PB effect in the forward case, as shown in Fig. \ref{fig4}(a). It results from the quantum interference between dominant excitation pathways and is consistent with the above analysed physical mechanism. In the backward case, the population of the first excited state is suppressed, corresponding to the minimum observed in Fig. \ref{fig2}(c). In contrast, the population of the second excited state is enhanced in the same detuning region, which is marked by the magenta five-pointed star in Fig. \ref{fig4}(a). This enhancement resembles the photon-induced tunneling effect; however, this feature does not constitute a local maximum. Specifically, the population of $|2\g\rangle$ is higher than that in the left detuning region but lower than that in the right region. The observed increase of population stems from the fact that the control laser effectively couples to relevant intermediate states at moderate atomic detunings. As the atomic detuning is further increased, however, the interaction between the atom and the control field weakens significantly. In this regime, the system approaches the standard driven Jaynes-Cummings model, and its dynamics become qualitatively similar to those of a cavity QED setup in the absence of auxiliary control.

\begin{figure}[b]
\centering
\includegraphics[width=0.75\columnwidth]{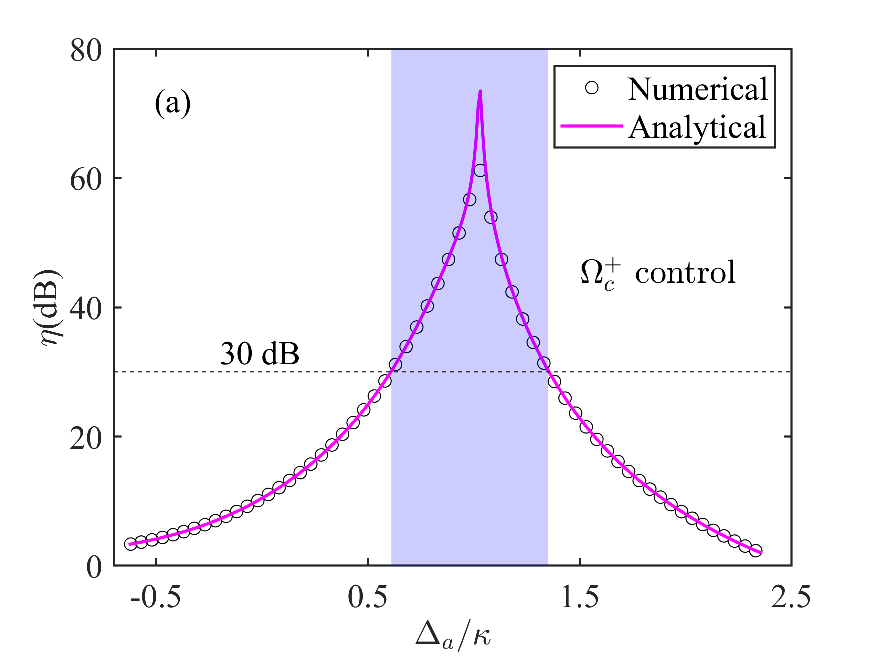}\\
\includegraphics[width=0.75\columnwidth]{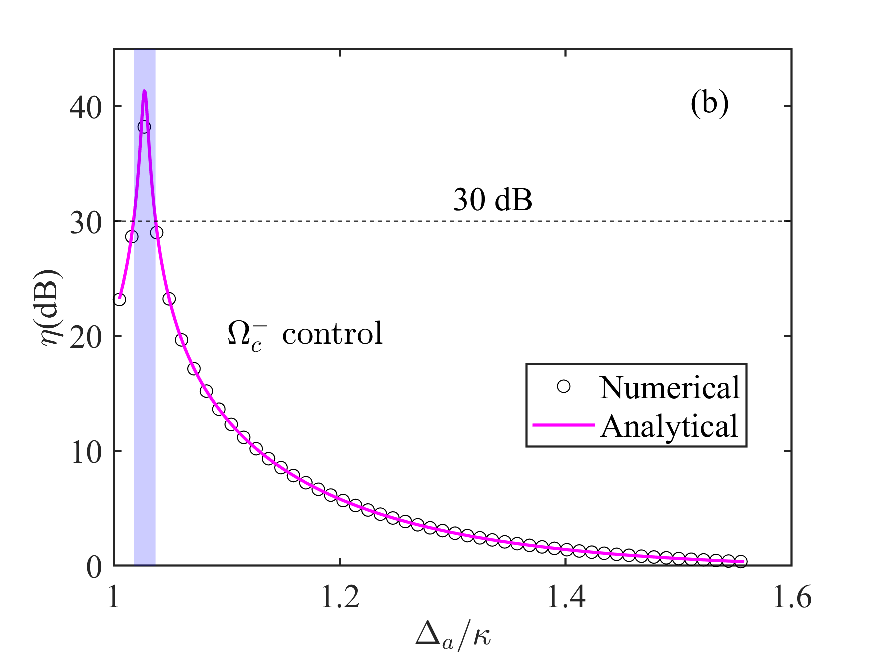}\\
\caption{The nonreciprocal ratio $\eta$ as a function of the atomic detuning $\Delta_a$. Panel (a) corresponds to the optimal control condition $\Omega_c^+$, and panel (b) to $\Omega_c^-$ control. The light-blue regions indicate that $\eta>30$ dB. The parameters are chosen those in Fig. \ref{fig2}.} \label{fig5}
\end{figure}

In Fig. \ref{fig4}(b), the optimal control condition $\Omega_c^-$ is applied. Although the population of the second excited state $|2\g\rangle$ remains very low for both forward and backward cases, the forward case exhibits a population that is approximately three orders of magnitude smaller than that of the backward case. This is attributed to the PB effect present in the forward case. Consistent with the previous analysis, one can verify that the population of the atomic excited state $|e\rangle$, obtained by summing over states $|0e\rangle$, $|1e\rangle$ and $|2e\rangle$, is significantly enhanced in this region. The increased occupancy of state $|0e\rangle$ enables a residual transition to state $|2\g\rangle$ via state $|1e\rangle$. In the forward case, the population of the single-photon state $|1\g\rangle$ is strongly suppressed, and it consequently reduces the population of state $|2\g\rangle$, placing the window of PB regime. However, in the backward case, the condition for suppressing the state $|1\g\rangle$ is not perfectly met by $\Omega_c^-$. This results in a direct excitation pathway  $|1\g\rangle\to|2\g\rangle$, which also fails to destructively interfere with other transition pathways, thereby leading to a non-blockaded state. This nonreciprocal PB feature is highlighted by the light-blue shaded region in Fig. \ref{fig2}(b).

\begin{figure}[b]
\centering
\includegraphics[width=0.75\columnwidth]{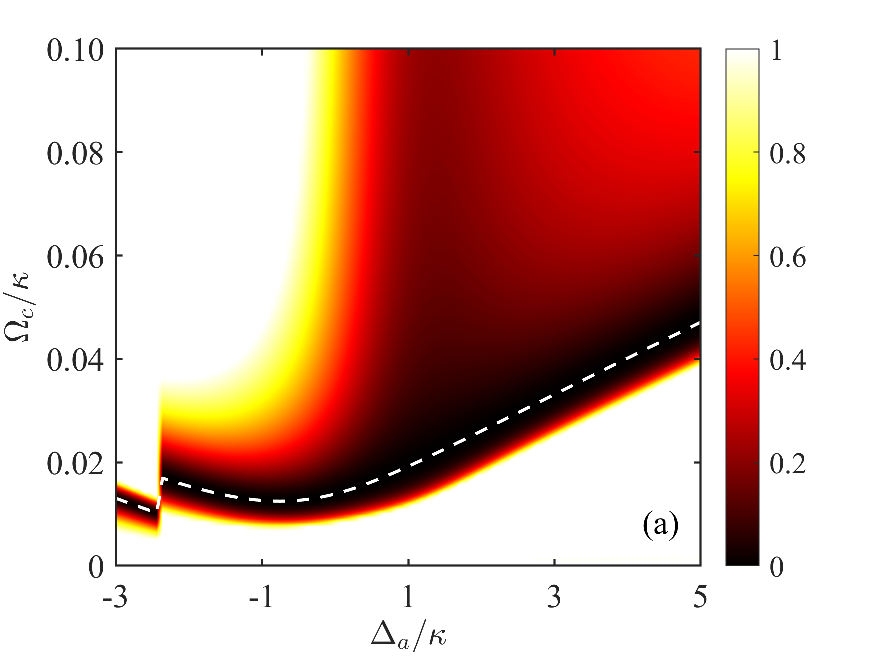}\\
\includegraphics[width=0.75\columnwidth]{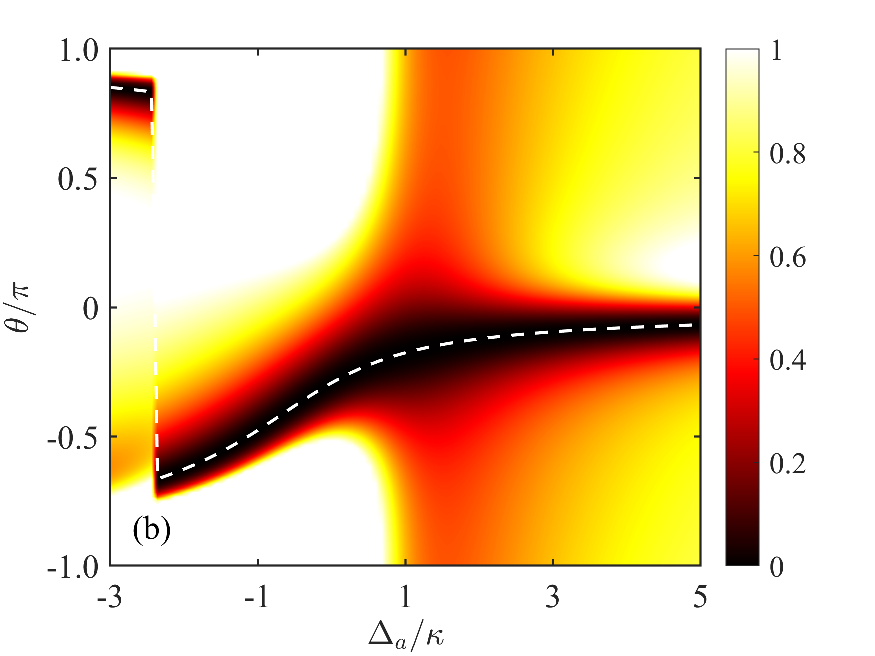}\\
\caption{The variation of the second-order correlation function $\g^{(2)}(0)$ on the atomic detuning $\Delta_a$ and the amplitude $\Omega_c$ of the control field in panel (a), and the relative phase $\theta$ in panel (b). The white dashed curves indicate the optimal control condition $\Omega_c^+$. The parameters are chosen those in Fig. \ref{fig2}.} \label{fig6}
\end{figure}

The nonreciprocal ratio of the PB effect between the forward and backward cases can be quantified by the rate of the corresponding second-order correlation functions, and be defined as
\begin{align}
\eta=-10\log_{10}\left[\frac{\g_f^{(2)}(0)}{\g_b^{(2)}(0)}\right],
\end{align}
where $\g_{f/b}^{(2)}(0)$ denotes the value of the second-order correlation function in the forward/backward propagating case. A large value of $\eta$ signifies a strong nonreciprocal PB effect, with effective blockade in the forward case and enhanced transmission in the backward case. In Fig. \ref{fig5}, we present the situation where the nonreciprocal ratio $\eta$ exceeds zero. The optimal control field $\Omega_c^+$ is employed in Fig. \ref{fig5}(a), while $\Omega_c^-$ is applied in Fig. \ref{fig5}(b). The magenta solid lines show the analytical result (\ref{Eq:g20}) derived via the probability amplitude method, and the black circles denote the numerical solutions obtained from the quantum master equation (\ref{Eq:master}). Excellent agreement is observed between the two approaches across the entire parameter range. The black dotted lines indicate the threshold of the nonreciprocal ratio $\eta=30$ dB, and the light-blue shaded regions indicate the parameter regimes where $\eta>30$ dB. According to the definition of the nonreciprocal ratio in Eq. (11), a high nonreciprocity requires the second-order correlation function in the forward configuration $\g^{(2)}_f(0)$ to be as small as possible, indicating a strong PB effect, while that in the backward case $\g^{(2)}_b(0)$ should be as large as possible. Based on the previous theoretical analysis, it is evident that both the magnitude of the nonreciprocal ratio and the associated bandwidth are significantly larger under $\Omega_c^+$ control compared to $\Omega_c^-$ control. This distinction is attributed to the enhancement of quantum destructive interference induced by $\Omega_c^+$.

\begin{figure}[t]
\centering
\includegraphics[width=0.75\columnwidth]{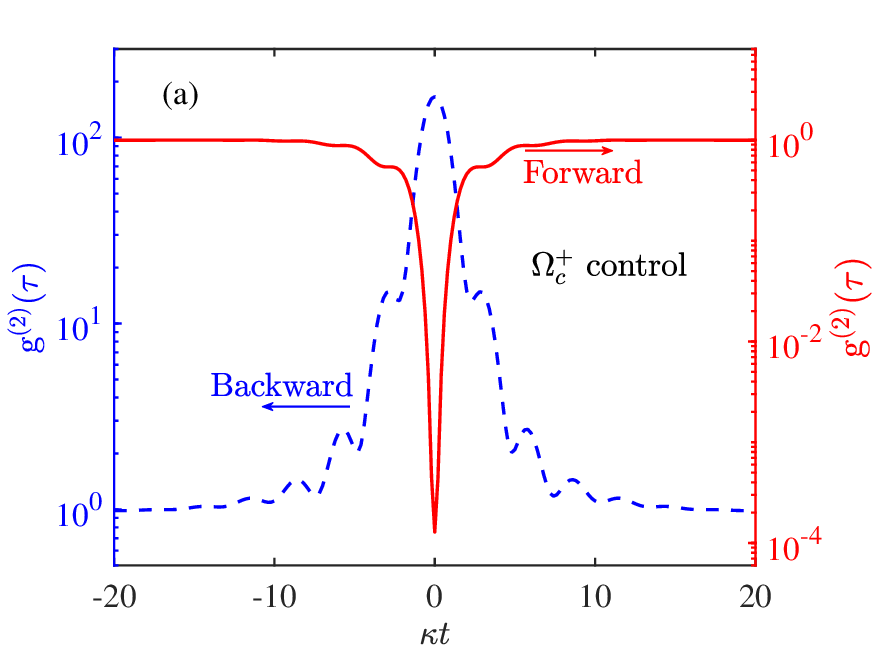}\\
\includegraphics[width=0.75\columnwidth]{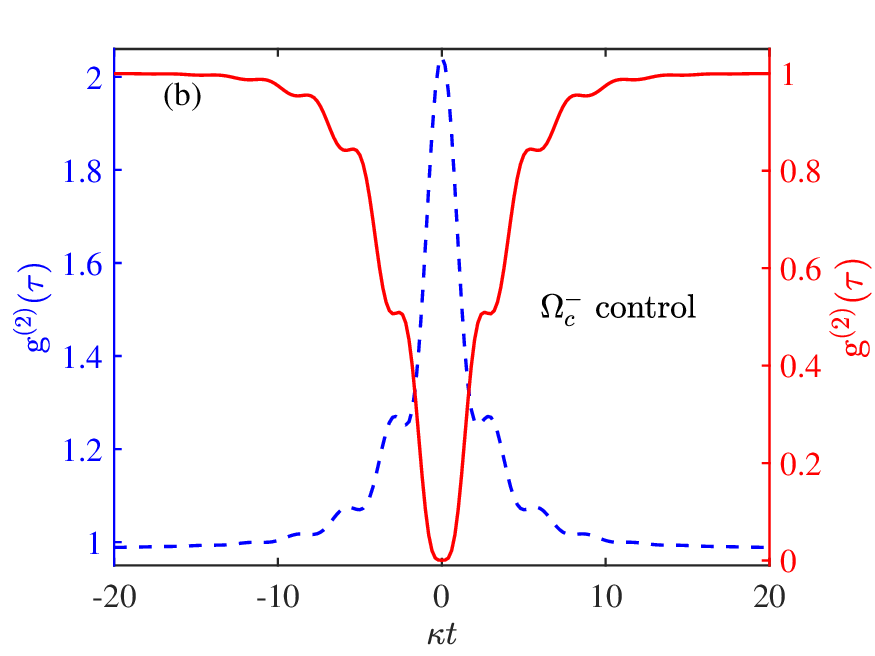}\\
\caption{The time-delayed second-order correlation function $\g^{(2)}(\tau)$ for the  forward (red solid) and backward (blue dashed) cases under the optimal control $\Omega_c^+$ in panel (a), and the optimal control $\Omega_c^-$ in panel (b). The parameters are set corresponding to the optimal PB points identified in Fig. \ref{fig2}.} \label{fig7}
\end{figure}

In Fig. \ref{fig6}(a), we show the variation of the second-order correlation function $\g^{(2)}(0)$ on the amplitude of the control field $\Omega_c$ and the atomic detuning $\Delta_a$. In Fig. \ref{fig6}(b), we display the dependence of the second-order correlation function $\g^{(2)}(0)$ on the relative phase $\theta$ between the control and driving lasers, as well as on the atomic detuning $\Delta_a$. The white dashed curves denote the expression of $\g^{(2)}(0)$ derived from the optimal control condition $\Omega_c^+$. For fixed values of other system parameters, the second-order correlation function $\g^{(2)}(0)$ reaches its minimum precisely along this optimal curve. An obvious change in the behavior of $\g^{(2)}(0)$ occurs near the atomic detuning $\Delta_a/\kappa\sim-2.43$, which originates from the properties of the complex square root function $\sqrt{\g^2\Delta_{ac}-\Omega^2\Delta_{ac}'}$ in $\Omega_c$. According to the theory of complex functions, the square root function is multi-valued and requires a branch cut, typically defined along the negative real axis $(-\infty,0]$, to be single-valued. As the parameter $\Delta_a$ varies, the imaginary part of the argument ${\g^2\Delta_{ac}-\Omega^2\Delta_{ac}'}$ approaches zero while the real part remains negative. Consequently, the argument approaches the branch cut. In this locus, the function exhibits singular behavior, which leads to a rapid variation in the function's value of the imaginary part. This further leads to a jump in the optimal control condition $\Omega_c^+$ in Eq. (\ref{Eq:+}), which is reflected in Fig. \ref{fig6}.

We further examine the time-delayed second-order correlation function $\g^{(2)}(\tau)$ under the control condition (\ref{Eq:optimal}). Fig. \ref{fig7}(a) shows $\g^{(2)}(\tau)$ evaluated with $\Omega_c^+$ control, while Fig. \ref{fig7}(b) displays the result under $\Omega_c^-$ control. In each panel, the red solid line corresponds to the forward case, while the blue dashed line represents the backward configuration. In the forward case, $\g^{(2)}(\tau)$ remains below 1 for all time with a pronounced minimum at $t=0$, it manifests the antibunching characteristic and confirms the presence of the PB effect. In contrast, the backward case consistently yields $\g^{(2)}(\tau)>1$, indicating a clearly bunched photon state. The parameters used in Fig. \ref{fig7} are chosen to match those at the perfect PB points identified in Fig. \ref{fig2}, ensuring a direct correspondence between the steady-state and temporal correlation analyses.

The approach proposed in Ref. \cite{SXWu25} involves embedding a second-order nonlinear crystal within an asymmetric Fabry-P\'{e}rot cavity, where the nonreciprocal PB effect is mediated by a parametric pumping. This strategy faces significant experimental challenges, as an efficient and stable parametric pumping is difficult even when the nonlinear crystal is placed in an external cavity; moreover, the pumping power must be carefully controlled: if it falls below the oscillation threshold, no down-converted field is generated, while an excessive pumping violates the weak-driving approximation under the PB regime. By comparison, this scheme is fully compatible with the current cavity QED setup, requiring only minor adjustments to laser parameters. A single coherent laser source can be split into driving and control beams, allowing for independent tuning of their amplitude ratio and relative phase $\theta$ via a tunable beam splitter and phase shifter. The time-delayed second-order correlation function of the output field can be directly measured using the established Hanbury Brown-Twiss or homodyne detection techniques. Through the input-output relation, the intracavity correlation properties are faithfully reflected in the transmitted photons. Consequently, the predicted nonreciprocal PB effect and its tunability via the control field can be experimentally verified with the current technology.

In contrast to the weak-driving regime, the few-photon ansatz breaks down under a strong driving, leading to significantly enhanced multiphoton processes. Consequently, the PB effect is weakened, and the nonreciprocal PB proposed in this model diminishes, causing the system to gradually exhibit classical behavior. Nevertheless, owing to the intrinsic asymmetry of the Fabry-P\'{e}rot cavity, the system remains a viable platform for exploring optical nonreciprocity \cite{PYang23}. In the presence of detuning between the driving and control fields, the system acts as a typical two-tone driven model. Therefore, the steady state becomes time-dependent, resulting in periodic modulation of the PB effect \cite{MLi22,YJing25}. This intriguing behavior motivates further exploration into the corresponding nonreciprocal PB effect.

\section{Conclusion}\label{sec4}
We propose a readily implementable scheme for realizing strong nonreciprocal photon blockade in a coherently driven atom-cavity system. By introducing an auxiliary control field with same frequency as the driving laser, we can engineer quantum interference among multiple excitation pathways, enabling directional suppression of two-photon states. Two distinct optimal control conditions, denoted by $\Omega_c^+$ and $\Omega_c^-$, emerge from the requirement of perfect photon blockade (e.g., $|C_{2\g}|=0$) within the few-photon subspace. Under $\Omega_c^+$ condition, the system exhibits a high-contrast nonreciprocal response with a ratio exceeding 30 dB and a broad blockade bandwidth; in contrast, $\Omega_c^-$ condition results in a markedly weaker nonreciprocity, which is attributed to the delicate quantum interference governed by the control field. Crucially, both the position and bandwidth of the photon blockade window can be dynamically tuned by adjusting the amplitude and relative phase of the control field, without requiring any structural modification to the device. Since this scheme relies only on standard coherent laser sources and second-order correlation measurements, it is fully compatible with current cavity QED platforms. This work offers a practical and versatile route toward nonreciprocal single-photon devices, with potential applications in quantum routing, quantum isolation, and integrated quantum information processing.

\section*{Acknowledgements}
This work was supported by National Natural Science Foundation of China (Grant No. 12204440), and Foundation of Shanxi Key Laboratory of Graphene Sensing Materials and Devices (No. SMX2025007).

\section*{Conflicts of Interest}
The author declares no conflicts of interest.

\section*{Data Availability Statement}
The data that support the findings of this study are available from the corresponding author upon reasonable request.

\end{document}